\documentclass[10pt,sigconf,letterpaper]{acmart}

\AtBeginDocument{%
  \providecommand\BibTeX{{%
    \normalfont B\kern-0.5em{\scshape i\kern-0.25em b}\kern-0.8em\TeX}}}

\usepackage{framed}
\usepackage{multirow}
\usepackage{booktabs}
\usepackage{ifthen}
\usepackage{color}
\usepackage{url}
\usepackage{amsmath}
\usepackage{xspace}
\usepackage[T1]{fontenc}
\usepackage{enumerate}
\usepackage[linesnumbered,ruled,vlined]{algorithm2e}
\usepackage{array,multirow,graphicx}
\usepackage{float}
\usepackage{balance}
\usepackage{tikz}
\usepackage{calc}
\usepackage{subfigure}
\usepackage{listings,amsfonts}
\usepackage{amsmath,xcolor,pifont}
\usepackage{url}
\usepackage{xspace}
\usepackage{hyperref,endnotes}
\usepackage{xcolor}
\usepackage{array}
\usepackage{multirow,makecell}
\usepackage[misc]{ifsym}

\definecolor{yellow}{RGB}{255,255,153}
\definecolor{grey}{RGB}{224,224,224}

\copyrightyear{2020}
\acmYear{2020}
\setcopyright{acmcopyright}\acmConference[ASEW '20]{35th IEEE/ACM International Conference on Automated Software Engineering Workshops}{September 21--25, 2020}{Virtual Event, Australia}
\acmBooktitle{35th IEEE/ACM International Conference on Automated Software Engineering Workshops (ASEW '20), September 21--25, 2020, Virtual Event, Australia}
\acmPrice{15.00}
\acmDOI{10.1145/3417113.3422186}
\acmISBN{978-1-4503-8128-4/20/09}

\begin{document}
\title{Market-level Analysis of Government-backed COVID-19 Contact Tracing Apps}

\author{Huiyi Wang}
\affiliation{%
  \institution{Beijing University of Posts and Telecommunications, China}
}

\author{Liu Wang}
\authornote{Co-first author.}
\affiliation{%
  \institution{Beijing University of Posts and Telecommunications, China}
}

\author{Haoyu Wang}

\affiliation{%
  \institution{Beijing University of Posts and Telecommunications, China}
}

\begin{abstract}
To help curb the spread of the COVID-19 pandemic, governments and public health authorities around the world have launched a number of contact-tracing apps. Although contact tracing apps have received extensive attentions from the research community, no existing work has characterized the users' adoption of contact tracing apps from the app market level. 
In this work, we perform the first market-level analysis of contact tracing apps. We perform a longitudinal empirical study (over 4 months) of eight government-backed COVID-19 contact tracing apps in iOS app store. We first collect all the daily meta information (e.g., app updates, app rating, app comments, etc.) of these contact tracing apps from their launch to 2020-07-31. Then we characterize them from release practice, app popularity, and mobile users' feedback. Our study reveals various issues related to contact tracing apps from the users' perspective, hoping to help improve the quality of contact tracing apps and thus achieving a high level of adoption in the population.
\end{abstract}

\maketitle

\section{Introduction}

The coronavirus pandemic (COVID-19) has quickly became a world-wide crisis. To help curb the spread of the COVID-19 pandemic, government and public health authorities around the world have adopted numerous measures, including lock-downs, social distancing, and contact tracing.
In general, \textit{contact tracing is the process of identification of persons who may have come into contact with an infected person (``contacts'') for identifying and supporting exposed individuals to be quarantined on time}~\cite{WikiContactTracing}. Past experience suggested that Contact tracing is the key to slowing the spread of contagious diseases, including COVID-19~\cite{CDCContactTracing}.

As a result, governments and public health authorities around the world have launched a number of contact-tracing mobile apps. By the time of this study, there are at least 50 contact tracing apps in both Google Play and iOS App Store. For example, \texttt{TraceTogether} was released on March 18, which is the first Bluetooth based COVID-19 contact tracing app released by the Singapore government. 
After that, a number of government-backed COVID-19 contact tracing smartphone apps have been launched, e.g., \texttt{AarogyaSetu} (India), \texttt{COVIDSafe} (Australia), \texttt{StopCovid France} (France), \texttt{Immuni} (Italy), \texttt{Corona Warn App} (Germany), etc. Besides, Apple and Google have collaborated to launch a framework that includes APIs and system-level supports to assist in enabling contact tracing.
Although different techniques (e.g., GPS and Bluetooth), architectures (e.g., centralized and decentralized) and protocols are used in these contact tracing apps, the general ideas are similar, i.e., by recording \textit{close proximity interactions} between individuals. Ideally, once found a new infected person, notifications can be sent to persons of a potential exposure to COVID-19, i.e., all the smartphones running the same contact tracing app that indicated the close proximity to the infected person.

Contact tracing apps have attracted extensive attentions from the research community. 
A number of papers have been focused on the privacy and security issues of the contact tracing apps~\cite{sun2020vetting, hobson2020trust}, the comparison of centralized and decentralized solutions~\cite{li2020decentralized}, new protocols~\cite{ZeroKnowledge}, etc.

\textit{Mobile user's adoption is the key to the success of contact tracing}. In short, to achieve accurate contact tracing and provide timely exposure notice, a high level of adoption in the population is a must. Although previous studies have revealed some obstacles (e.g., privacy concern) leading to the ineffectiveness of contact tracing, no existing work has systematically characterized the user's adoption from market level. App markets allow users to leave their ratings and comments after downloading and using each app~\cite{wang2018beyond}. These comments can be considered as the direct feedback from users who have experienced the apps, which can help developers address the issues. However, to the best of our knowledge, existing research papers have not studied the user comments of contact tracing apps in details.

\begin{table*}[t!]
\caption{The 8 government-backed COVID-19 contact tracing apps studied in this paper.}
\vspace{-0.15in}
\label{tab:covid19apps}
\resizebox{1\linewidth}{!}{
\begin{tabular}{@{}ccccccc@{}}
\toprule
 Country & App Name & Category & Bundle ID & Developer & Launch Time & Compatibility\\ \midrule
Japan & COCOA & Medical & jp.go.mhlw.covid19radar & Ministry of Health, Labour and Welfare - Japan & 2020-06-18 & $\geq$ iOS 13.5\\ 
Germany & Corona-Warn-App & Health\& Fitness & de.rki.coronawarnapp & Robert Koch-Institut & 2020-06-15 & $\geq$ iOS 13.5 \\ 
France & StopCovid France & Health \& Fitness & fr.gouv.stopcovid.ios & Gouvernement Francais & 2020-06-02 & $\geq$ iOS 11.4\\ 
Italy & Immuni & Medical & it.ministerodellasalute.immuni & Ministero della Salute & 2020-06-01 & $\geq$ iOS 13.0 \\ 
Australia & COVIDSafe & Health \& Fitness & au.gov.health.covidsafe & Department of Health, Australian Capital Territory & 2020-04-25 & $\geq$ iOS 10.0 \\ 
India &  AarogyaSetu & Health \& Fitness & in.nic.arogyaSetu & NATIONAL INFORMATICS CENTRE & 2020-04-01 & $\geq$ iOS 10.3 \\ 
Austria & Stopp Corona & Medical & at.roteskreuz.stopcorona & Osterreichisches Rotes Kreuz & 2020-03-27 & $\geq$ iOS 13.5 \\ 
Singapore & TraceTogether & Medical & sg.gov.tech.bluetrace & Government Technology Agency & 2020-03-18 & $\geq$ iOS 10.0 \\ 
\bottomrule
\end{tabular}
}
\vspace{-0.1in}
\end{table*}

\textbf{This Paper.}
In this work, we perform the market-level analysis of contact tracing apps, in order to understanding mobile users' adoption of them. To be specific, we focus on eight government-backed COVID-19 contact tracing apps in the iOS App Store. 
We perform a longitudinal study of these apps from the perspective of app market, by first collecting all the daily information of these apps (see \textbf{Section~\ref{subsec:data}}), including app updates, app ranking, app rating, and user comments, from the day of their launch to July 31, 2020. 
We then characterize them from 
1) \textit{release practice} (see \textbf{Section~\ref{subsec:release}}), including the release time and release frequency, especially correlating with the trends of the COVID-19 (i.e., infected people) in the corresponding country; 
2) \textit{app popularity} (see \textbf{Section~\ref{subsec:popularity}}), i.e., the rankings of these contact tracing apps across countries, which can reflect users' adoption to some extent; and
3) \textit{mobile users' feedback} (see \textbf{Section~\ref{subsec:user}}), including app ratings and user comments.

Our work revealed that although it shows the promising beginning of contact tracing apps across the eight countries we studied, there remain a number of emerging obstacles leading to the ineffectiveness of contact tracing, including \textit{Bluetooth Conflict}, \textit{Compatibility Issue}, \textit{Energy Issue}, \textit{Privacy Issue}, \textit{Accuracy and Usability issues}, etc. We hope this paper can serve as indicators from mobile users to help improve the quality of contact tracing apps, thus enhancing the adoption rate and effectiveness of contact tracing.

\vspace{-0.1in}
\section{Related Work}
\label{sec:related}

A number of studies~\cite{sun2020vetting, hobson2020trust, li2020decentralized} have analyzed the contact tracing apps, mainly from the perspective of security and privacy. 
Li et al.~\cite{li2020decentralized} presented a survey study to analyze the user's perceptions of the utility of contact tracing apps and the potential privacy risks due to the collection and releasing of sensitive user data. Sun et al.~\cite{sun2020vetting} analyzed the security and privacy of 34 contact tracing apps, including potential vulnerabilities, privacy leaks, and the robustness of privacy protection approaches. 
Besides, new solutions and protocols~\cite{ZeroKnowledge} are proposed to enhance existing contact tracing techniques.
Beyond these studies, a few studies have performed general analysis of COVID-19 themed apps and threats. 
For example, He et al.~\cite{he2020virus} is the first to study COVID-19 themed mobile malware. Xia et al.~\cite{xia2020dont} characterized COVID-19 themed cryptocurrency scams. Although several research proposed to study users' concerns of contact tracing apps~\cite{redmiles2020user, li2020decentralized}, none of existing approaches have systematically analyzed the mobile users' feedback in app markets.

\section{Study Design}
\label{sec:design}

\subsection{Research Questions}

Our study is driven by the following research questions:

\begin{itemize}
    \item[RQ1] \textit{What is the release practice of contact tracing apps?} We seek to study: 1) when are these contact tracing apps released? Especially when correlating with the trend of the COVID-19 in corresponding countries; and 2) How frequently do contact tracing apps update? As contact tracing is the key to slowing the spread of COVID-19, we are wondering whether the apps evolve frequently to address bugs and demanding issues.
    \item[RQ2] \textit{What is the popularity of contact tracing apps?} 
    We want to study the popularity (i.e., the app ranking and app downloads) of contact tracing apps, which can reflect the mobile users' early adoption of these apps.
    \item[RQ3] \textit{What are users' concerns of contact tracing apps?} 
    Use ratings and comments can be considered as the direct feedback from users who have experienced contact tracing apps. Thus, we want to study users' concerns from their comments.
\end{itemize}

\subsection{Data Collection}
\label{subsec:data}

As aforementioned, there are dozens of contact tracing apps in the app market. 
In this paper, we consider 8 government-backed contact tracing apps, as they are the most widely used COVID-19 apps in the corresponding countries. In this way, we can gain the overall understanding of the users' concerns of contact tracing in the country. For the app market, we select iOS App Store, one of the most popular app markets. The basic information of the selected 8 apps is shown in Table~\ref{tab:covid19apps}. By the time of this study, these eight apps are the only available government-backed contact tracing apps in iOS app store.
We have collected the \textit{daily} market-level information of them, include app ranking (in both its category ranking and the overall ranking), app updates, app ratings, and all the app comments, from their launch to 2020-07-31.

\begin{figure*}[t]
\centering
  \includegraphics[width=0.9\textwidth]{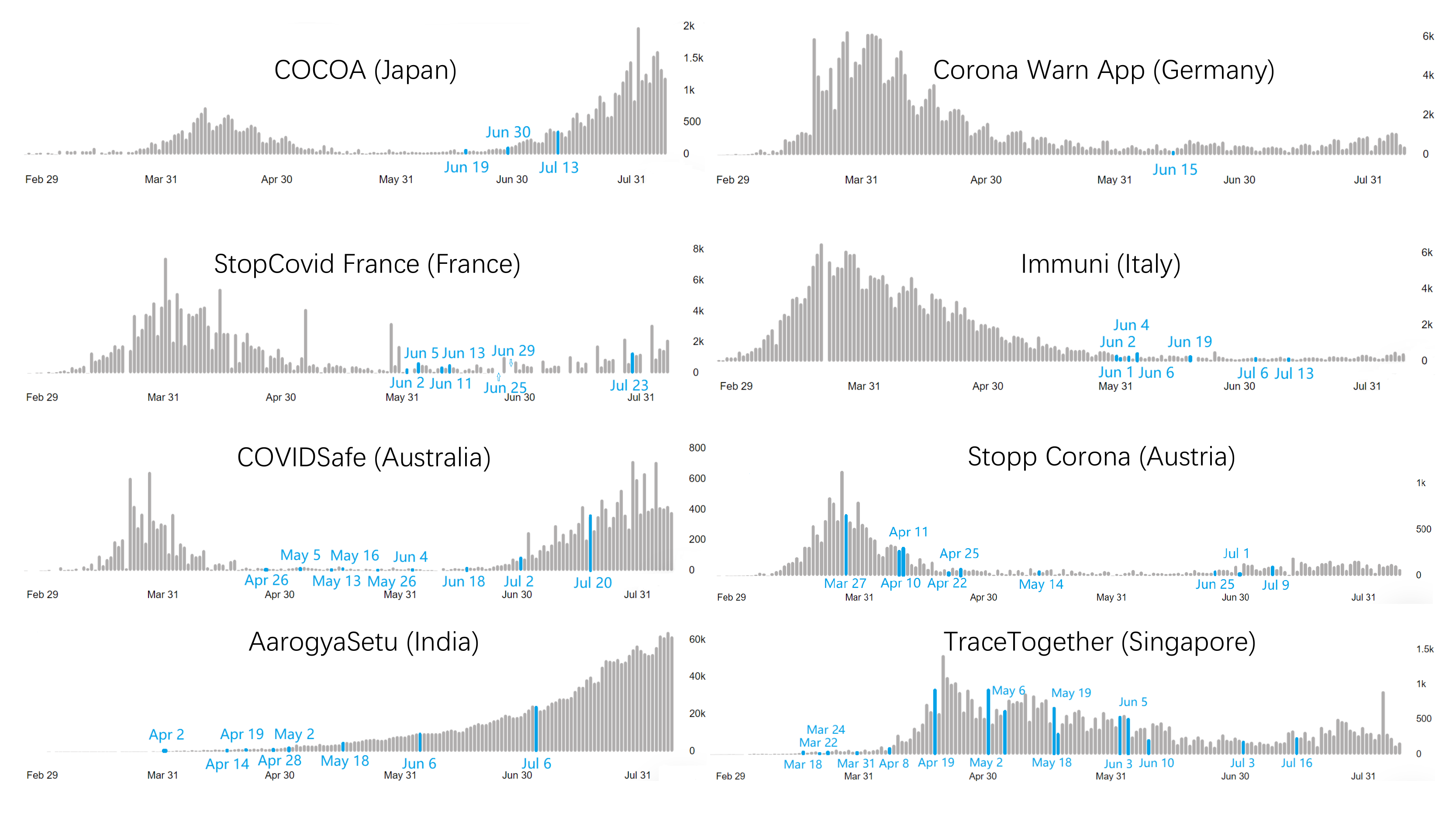}
  \vspace{-0.2in}
\caption{The Release Time of Contact Tracing Apps across Different Countries. The Y-axis suggests the newly reported infected COVID-19 cases in each country (data from https://covid19.who.int/). The blue columns represent the launch time/update time of the corresponding contact tracing app.}
  \vspace{-0.1in}
\label{fig:contactTracingTime} 
\end{figure*}

\section{Empirical Study}
\label{sec:empirical}

\subsection{RQ1: Release Practice}
\label{subsec:release}

\subsubsection{Release Time of Contact Tracing Apps}

As shown in Figure~\ref{fig:contactTracingTime}, \texttt{TraceTogether} is the first contact tracing app launched by the Singapore government at March 18, 2020. After that, \texttt{Stopp Corona} was launched by the Austria government at March 27 and \texttt{AarogyaSetu} was released by the India government at April 2.
It is obvious to see that, Singapore and India have released their contact tracing apps before the break of COVID-19 in their countries, and Austria launches the contact tracing app during the peak of the COVID-19 infection in the country. While for the remaining studied countries (Japan, France, Australia, Germany, Italy), their government-backed contact tracing apps are only available after the first first wave of the COVID-19 pandemic in their countries.
\textit{The release time can reflect that, for most countries, the contact tracing apps have not been timely launched to help curb the spread of COVID-19 pandemic.}

\begin{figure}[t]
\centering
  \includegraphics[width=0.43\textwidth]{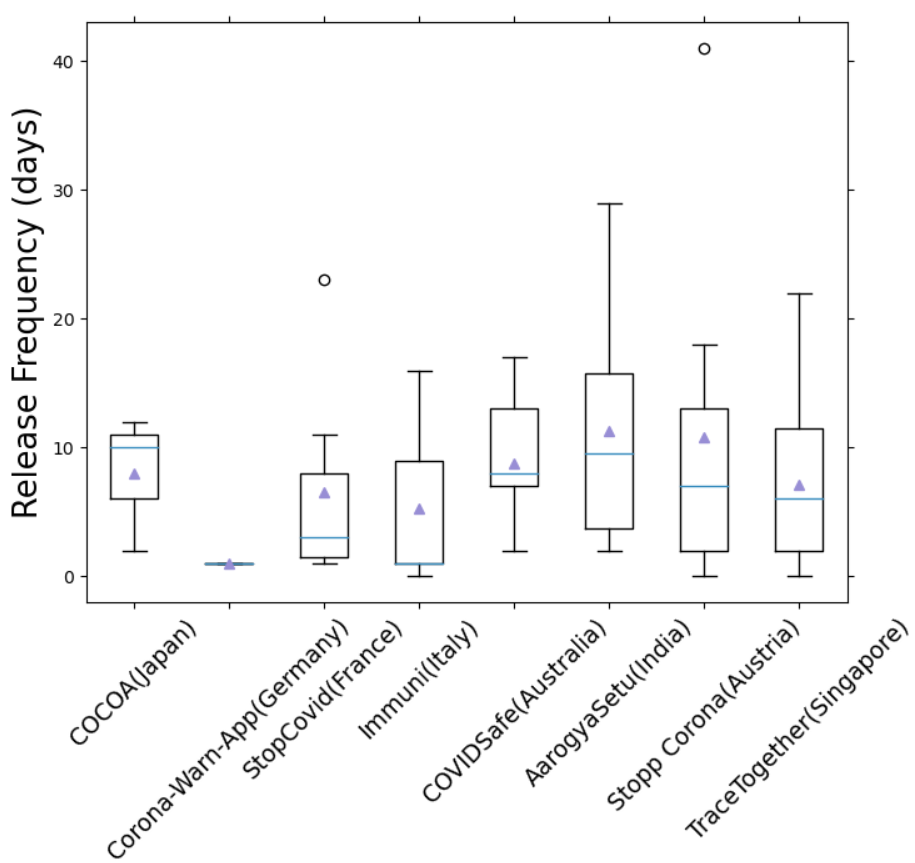}
  \vspace{-0.1in}
\caption{Release Frequency of Contact Tracing Apps.}
  \vspace{-0.2in}
\label{fig:releasefrequency} 
\end{figure}

\subsubsection{Release Frequency of Contact Tracing Apps}

We further study the release frequency of contact tracing apps. As shown in Figure~\ref{fig:releasefrequency}, most of contact tracing apps update frequently, i.e., the median time between two releases is less than 10 days in general. From Figure~\ref{fig:contactTracingTime} we can further observe that, during the early stage after the launch of contact tracing apps, they usually update in a rapid manner (e.g., 2 or 3 days to update a new version). As a comparison, Wang et al.~\cite{GPlayEvolution} found that more than 60\% apps did not release new versions for almost four years during their study.
Thus, \textit{contact tracing apps are usually updated frequently in general}.

\subsection{RQ2: App Popularity}
\label{subsec:popularity}

\begin{figure}[t]
\centering
  \includegraphics[width=0.4\textwidth]{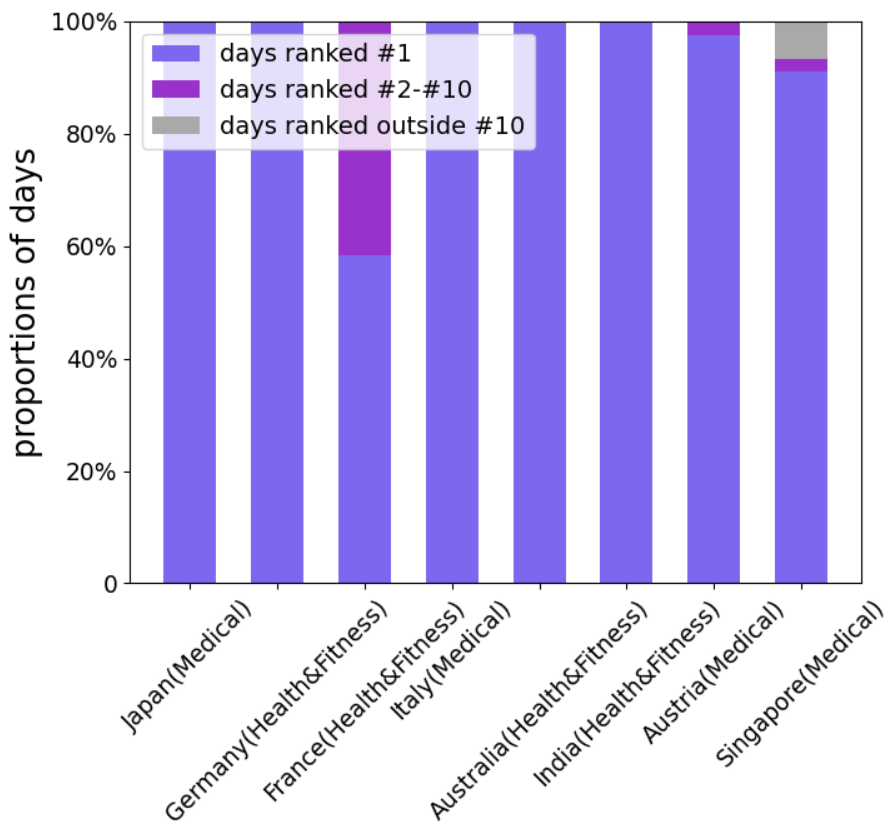}
    \vspace{-0.2in}
\caption{Ranking of Contact Tracing Apps in the Corresponding Categories.}
  \vspace{-0.2in}
\label{fig:appcategoryranking} 
\end{figure}

\begin{figure}[t]
\centering
  \includegraphics[width=0.4\textwidth]{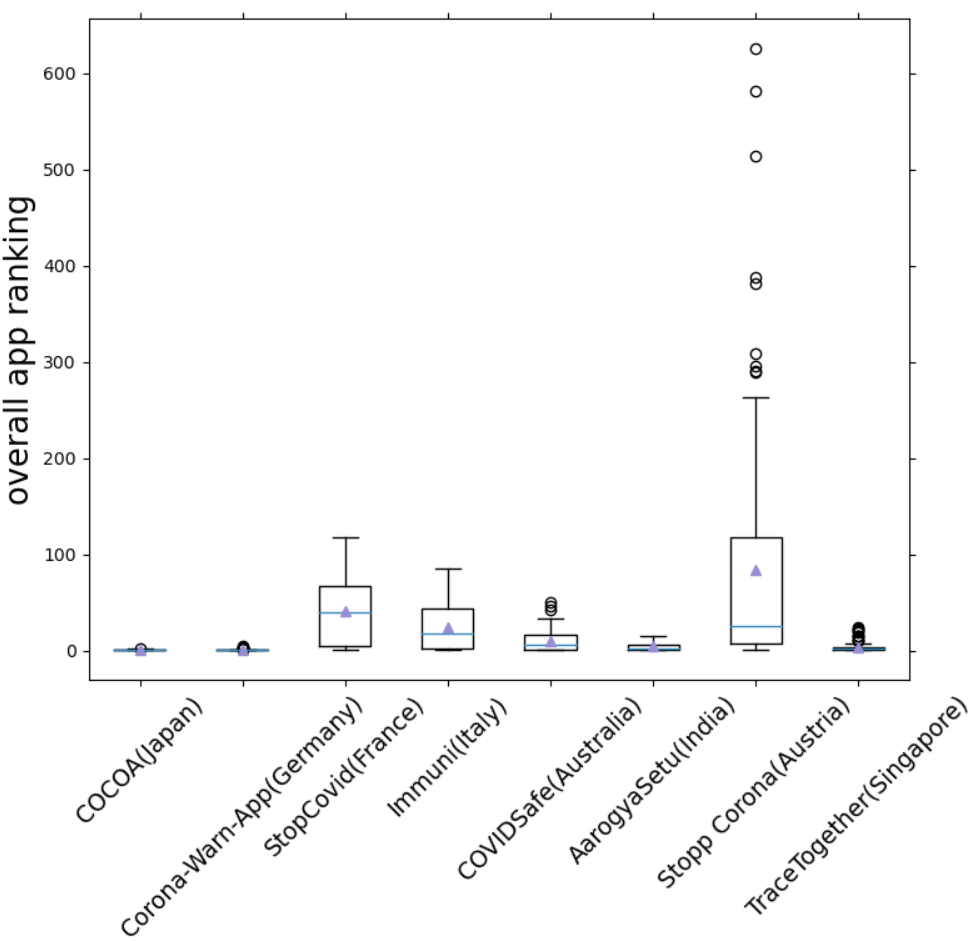}
  \vspace{-0.2in}
\caption{Overall Ranking of Contact Tracing Apps across Time (from launch to 2020-07-31).}
  \vspace{-0.1in}
\label{fig:overallranking} 
\end{figure}

Next, we study app popularity. As we cannot get the number of app installs from iOS App Store, we refer to the daily app ranking as the indicator of app popularity. We have studied two kinds of app rankings: 1) app ranking in its category and 2) the overall app ranking in iOS app market. 

As shown in Figure~\ref{fig:appcategoryranking}, most of contact tracing apps remain the top-1 in their categories (Medical or Health \& Fitness). Among them, five apps (\texttt{COCOA}, \texttt{Corona-Warn-App}, \texttt{Immuni}, \texttt{COVIDSafe}, and \texttt{Aarogyasetu}) always rank first since their launch. As to the overall app ranking, most of the apps can rank top-100 or even top-10 (e.g., \texttt{COCOA} and \texttt{Corona-Warn-App}) all the time. The only exception is \texttt{Stopp Corona} in Austria, i.e., its ranking fluctuates greatly. It ranks beyond 500 for some days. Nevertheless, \textit{the app ranking shows the promising beginning of contact tracing apps}.

\subsection{RQ3: Mobile Users' Feedback}
\label{subsec:user}

Then we study mobile users' feedback of these contact tracing apps, including \textit{app ratings} and \textit{user comments}.

\subsubsection{App Ratings}
\begin{figure}[t]
\centering
  \includegraphics[width=0.4\textwidth]{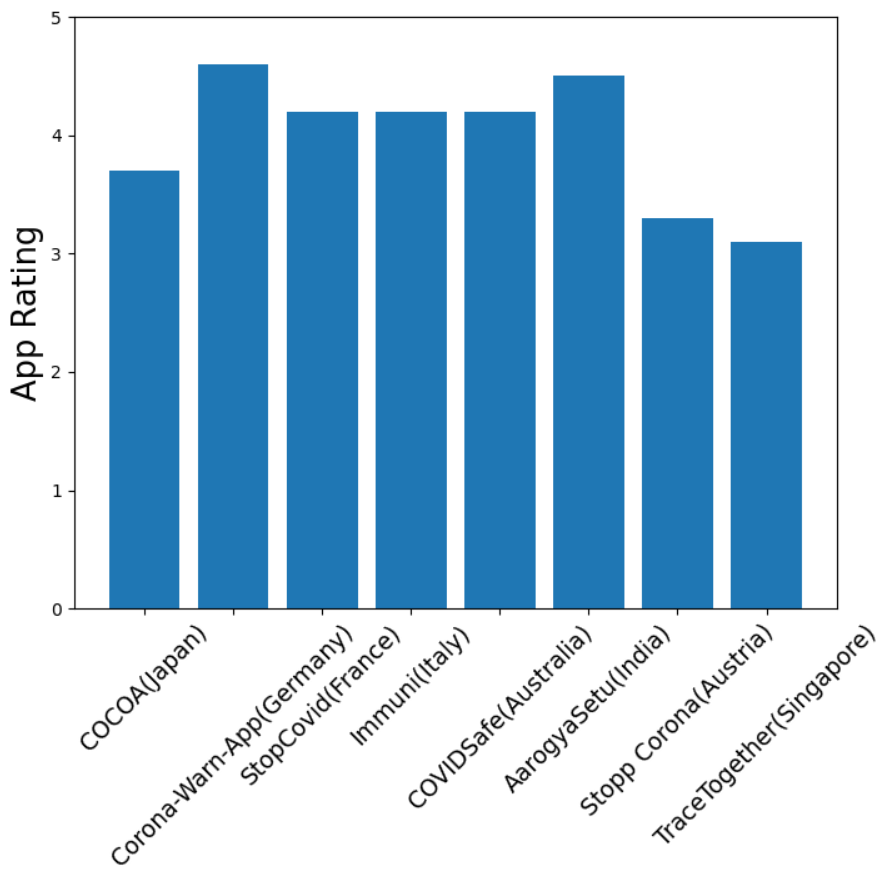}
\vspace{-0.1in}
\caption{App Ratings of Contact Tracing Apps.}
\vspace{-0.1in}
\label{fig:apprating} 
\end{figure}

As shown in Figure~\ref{fig:apprating}, most apps receive high app ratings, i.e., 5 of them have received ratings higher than 4. It suggests that most users are positive towards these contact tracing apps. 
\texttt{Corona-Warn-App} receives the highest rating (4.6).
\texttt{TraceTogether} has the lowest app rating (3.1). 

\begin{table}[h]
\caption{The Collected User Comments.}
\vspace{-0.1in}
\label{tab:comments}
\resizebox{\linewidth}{!}{
\begin{tabular}{@{}cccc@{}}
\toprule
 Country & App Name & \#Comments & \#1-star Comments (\%)\\ \midrule
 Japan & COCOA & 1,950 & 720 (36.9\%) \\
 Germany & Corona-Warn-App & 8,983 & 2,259 (25.1\%) \\
 France & StopCovid France & 850 & 319 (37.5\%) \\
 Italy & Immuni & 3,107 & 940 (30.3\%) \\
 Australia & COVIDSafe & 3,402 & 1,200 (35.3\%) \\
 India & AarogyaSetu & 3,691 & 1,260 (34.1\%) \\
 Austria & Stopp Corona & 499 & 300 (60.1\%) \\
 Singapore & TraceTogether & 775 & 360 (46.5\%) \\ \hline
  & Overall & 23,257  & 7,358 (31,6\%)  \\

\bottomrule
\end{tabular}
}
\vspace{-0.1in}
\end{table}

\subsubsection{Taxonomy of User Comments}
We further analyze user comments to understand their concerns. As shown in Table~\ref{tab:comments}, we have collected 23,257 user comments in total\footnote{We collect all the non-empty comments. As some users only rate the app without comments, the number might be different with that of app rating.}. To investigate user concerns, we focus only on the 1-star comments (7,358), which are in general mainly user complaints. To make an accurate classification of these comments, the first two authors examined 1,000 randomly selected comments and summarized the following 9 categories of complaints (including ``others'' category). 
Then, based on this taxonomy, the first two authors manually label the category of each 1-star comments independently. For the disagreements of category labelling, a further discussion is performed. 
Note that there are multiple languages in the comments, e.g., Japanese for \texttt{COCOA}, and Italian for \texttt{Immuni}. Thus, we take advantage of Google Translation to convert all the comments into English before manually labelling.
It takes two people three (3) days to label all the 7,358 1-star comments. 
\textit{We know this process might not be scalable. However, our sole purpose is to accurately understand users' concerns and measure the proportion of these concerns, while this is the most accurate approach.}

We next describe each category and will further measure their proportion in Section~\ref{subsec:commentmeasure}. 

1) \textit{\textbf{Bluetooth Conflict}}. Bluetooth techniques are widely used as the proximity sensing method to identify close proximity interactions between individuals. However, it may introduce Bluetooth conflict issue. We have identified hundreds of complaints. For example, some users complain that it breaks Bluetooth connections to Apple Watch (see Comment Example \#1). This issue leads to the uninstallaton of contact tracing apps for many people.

\begin{framed}
\noindent \textbf{Example \#1 Bluetooth Conflict (from COVIDSafe):} 
\textit{
``The app is a good idea,  but it breaks Bluetooth connections to other devices, Apple Watch in particular. I had to delete the app because of this. I had a friend tell me they were going to take their Apple Watch in for a battery replacement until I asked if they were running COVIDSafe. Once they deleted the app their problems went away.'' 
}
\end{framed}

2) \textit{\textbf{Compatibility Issue}}.
To achieve a high level of adoption in the population, the contact tracing app should have good compatibility, i.e., supporting most devices. However, we find that the compatibility issue is a serious problem in contact tracing apps. As summarized in Table~\ref{tab:covid19apps}, many of them support only high level iOS versions. Three contact tracing apps require iOS 13.5 or later. Many users complain about it (see Example \# 2). We think it might be a great obstacle leading to the ineffectiveness of contact tracing.

\begin{framed}
\noindent \textbf{Example \#2 Compatibility Issue (from Corona-Warn-App):} 
\textit{
``The app cannot be loaded on an iPhone 6 with the current software (12.4.7), there is no more up-to-date software. What a pity!''
}
\end{framed}

3) \textit{\textbf{Google/Apple Framework Support}}.
As aforementioned, Google and Apple have launched a comprehensive Bluetooth based framework that includes APIs and system-level techniques to enable contact tracing. However, by the time of this study, only four apps (COCOA, Immuni, Stopp Corona and Corona-Warn-App) have supported such contact tracing framework. Thus, some users complained that contact tracing apps should support this unified framework, otherwise the app is useless (see Comment Example \# 3). Besides, without using this system-level supported framework, the energy/battery issue might be serious.

\begin{framed}
\noindent \textbf{Example \#3 Framework (from COVIDSafe):} 
\textit{
``iOS 13.5 includes a contact tracing API. This app in its current form is useless, unless it takes advantage of the new API.''
}
\end{framed}

\begin{figure}[t]
\centering
  \includegraphics[width=0.5\textwidth]{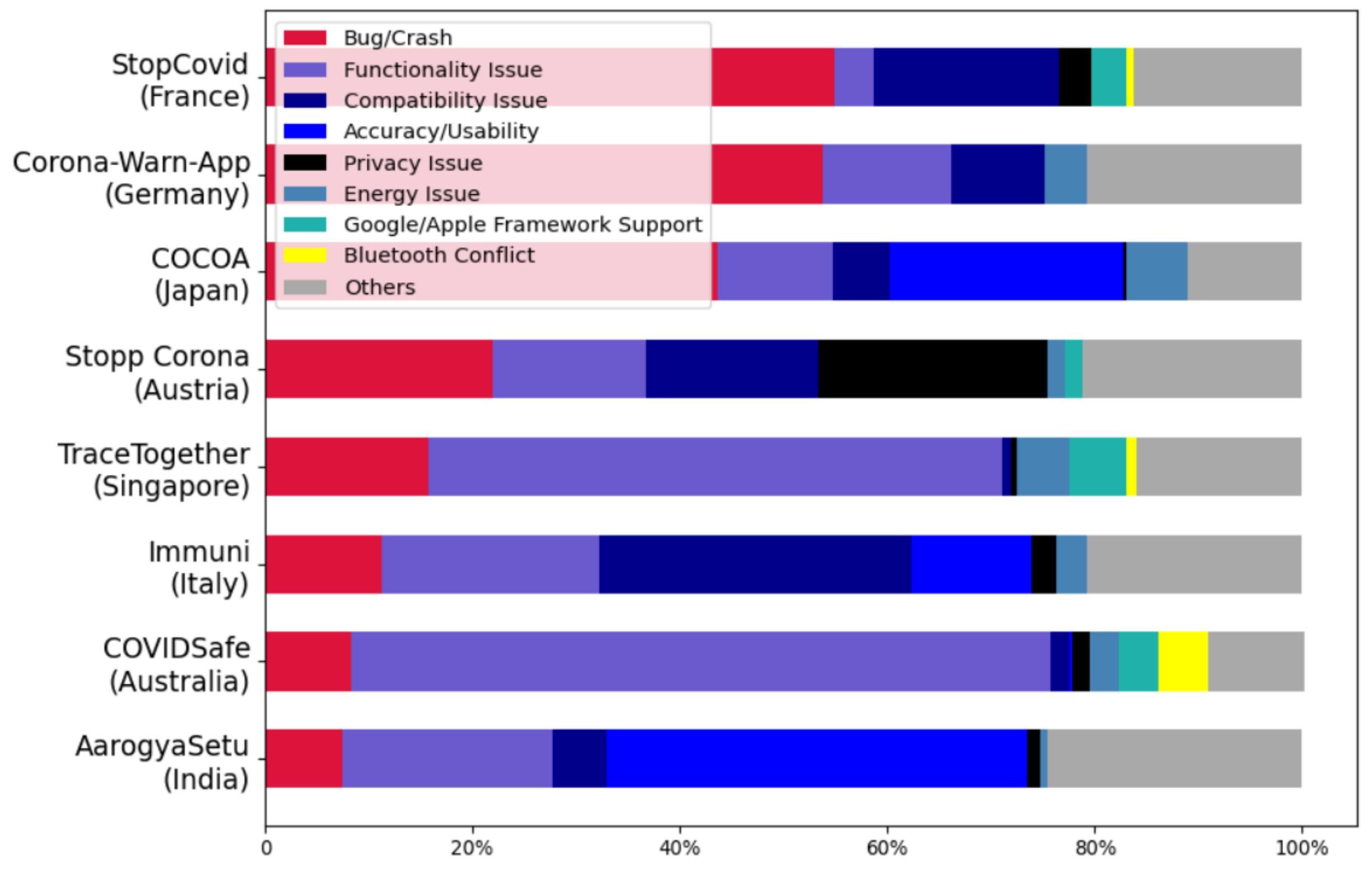}
  \vspace{-0.25in}
\caption{Distribution of User Complaints across Eight Contact Tracing Apps.}
  \vspace{-0.2in}
\label{fig:commentdistribution} 
\end{figure}

4) \textit{\textbf{Energy Issue}}.
As the contact tracing apps should continuous record the close proximity interactions between individuals, the energy issue is a big problem. It is no doubt that many users complain it (see Comment Example \#4). It might be annoying to many people that they have to uninstall the contact tracing app. 

\begin{framed}
\noindent \textbf{Example \#4 Energy Issue (from TraceTogether):} 
\textit{
``It wasting my iPhone battery life is draining fast when I turn on Bluetooth and do nothing at all.''
}
\end{framed}

5) \textit{\textbf{Privacy Issue}}.
Privacy remains a obstacle in the contact tracing app. Although many previous work discussed the privacy issue and proposed various privacy-preserving approaches, we are unaware users' concerns. We indeed find privacy complaints in the apps, however, the proportion is not high. We will discuss it in the following. 

\begin{framed}
\noindent \textbf{Example \#5 Privacy Issue (from AarogyaSetu):} 
\textit{
``This app gathers your location 24/7 and even through Bluetooth. Please note that by signing up on this app you will be giving out your personal information and location of all times to the developer.''
}
\end{framed}

6) \textit{\textbf{Bug/Crash}}.
Although bug and crash can be observed in any app, the impact to contact tracing might be more serious. It directly stops users from using the app. The following is an example comment from COVIDSafe.

\begin{framed}
\noindent \textbf{Example \#6 Crash (from COVIDSafe):} 
\textit{
``Have tried to register twice but it keeps crashing.''
}
\end{framed}

7) \textit{\textbf{Functionality Issue}}.
We further observe a large portion of functionality issues. For example, many users of COVIDSafe complain that only when the app is running in the foreground can it work (see Comment Example \#7).

\begin{framed}
\noindent \textbf{Example \#7 Functionality Issue (from COVIDSafe):} 
\textit{
``Endless notifications reminding me to use it - but none of them mention that the app needs be be running in the foreground of an unlocked phone to work.''
}
\end{framed}

8) \textit{\textbf{Accuracy/Usability}}.
Many users question the accuracy and usability. As we all know, the effectiveness of the contact tracing relies on the high percentage adoption of users. However, although these apps are quite popular in the app markets, the adoption rate is far less than adequate. Thus, some users raise such concerns (see Comment Example \#8).

\begin{framed}
\noindent \textbf{Example \#8 Accuracy/Usability (from COCOA):} 
\textit{
``For more than 6 million downloads, I don't think I can confirm contact with three people who are positive.''
}
\end{framed}

9) \textit{\textbf{Others}}.
Besides, a number of comments we cannot classify them into the above categories. Some of are too short (e.g., less than 5 words), thus we cannot extract the semantic information. Some of them are related to the trust or political concerns, e.g., ``\textit{Do you trust the Japanese government?}''.

\subsubsection{Distribution of User Complaints}
\label{subsec:commentmeasure}
We further measure the distribution of different kinds of user complaints, as shown in Figure~\ref{fig:commentdistribution}. It is interesting to see that, the distribution of user concerns varies greatly across apps (countries). 
Bug and crash issues are prevalent in \texttt{StopCovid} (54.9\%) and \texttt{Corona-Warn-App} (53.8\%), while they only account for a small proportion in \texttt{AarogyaSetu} (7.5\%) and \texttt{COVIDSafe} (8.3\%). Bluetooth Conflict complaints are only found in \texttt{COVID\\Safe} (4.8\%), \texttt{TraceTogether} (1.1\%) and \texttt{StopCovid} (0.6\%). Mobile users of \texttt{Stopp Corona} care most of privacy issues, with 22\% of complaints are related to privacy. Accuracy and Usability concerns are mostly found in \texttt{AarogyaSetu} (40.6\%). Functionality issues are dominant in
\texttt{TraceTogether} (55.3\%) and \texttt{COVIDSafe} (67.4\%), while they only account for 3.8\% in \texttt{StopCovid}.
\textit{This result suggests that there are many demanding issues waiting to be addresses in contact tracing apps, which may lead to the ineffectiveness of contact tracing.}


\section{Conclusion}

In this paper, we study the government-backed COVID-19 contact tracing apps from the perspective of app markets. Our longitudinal study includes app popularity analysis, app popularity analysis and user feedback characterization. Although these contact tracing apps show early promising ranking in app market, there are a number of unsolved emerging issues. 
We believe that our research efforts can positively contribute to the development of contact tracing apps, and boost the focus of related topics for the research community.

\section*{Acknowledgment}
This work was supported by the National Natural Science Foundation of China (No. 61702045).
Haoyu Wang is the corresponding author (haoyuwang@bupt.edu.cn).

\bibliographystyle{plain}
\bibliography{cite}

\end{document}